\documentclass[runningheads]{llncs}
\usepackage{graphicx}
\usepackage{url}
\newcommand{\ia}{\emph{Internet Archive}}
\newcommand{\msc}{\emph{MS-Celeb-1M}}
\usepackage{multirow}
\usepackage{array}

\newcolumntype{L}[1]{>{\raggedright\let\newline\\\arraybackslash\hspace{0pt}}m{#1}}
\newcolumntype{C}[1]{>{\centering\let\newline\\\arraybackslash\hspace{0pt}}m{#1}}

\begin{document}
\title{Finding Person Relations in Image Data of the Internet Archive}
\titlerunning{Finding Person Relations in Image Data of the Internet Archive}

\author{Eric M\"uller-Budack$^{1,2}$\orcidID{0000-0002-6802-1241} \and \\Kader Pustu-Iren$^1$\orcidID{0000-0003-2891-9783} \and Sebastian Diering$^1$ \and \\Ralph Ewerth$^{1,2}$\orcidID{0000-0003-0918-6297}}
\authorrunning{M\"uller-Budack et al.}
\institute{Leibniz Information Centre for Science and Technology (TIB), Hannover, Germany \and L3S Research Center, Leibniz Universit\"at Hannover, Germany}

\maketitle    
\begin{abstract}
The multimedia content in the World Wide Web is rapidly growing and contains valuable information for many applications in different domains. For this reason, the Internet Archive initiative has been gathering billions of time-versioned web pages since the mid-nineties. 
However, the huge amount of data is rarely labeled with appropriate metadata and automatic approaches are required to enable semantic search.
Normally, the textual content of the Internet Archive is used to extract entities and their possible relations across domains such as politics and entertainment, whereas image and video content is usually neglected. In this paper, we introduce a system for person recognition in image content of web news stored in the Internet Archive. Thus, the system complements entity recognition in text and allows researchers and analysts to track media coverage and relations of persons more precisely. 
Based on a deep learning face recognition approach, we suggest a system that 
automatically detects persons of interest and gathers sample material, which 
is subsequently used to identify them in the image data of the Internet 
Archive. We evaluate the performance of the face recognition system on an
appropriate standard benchmark dataset and demonstrate the feasibility of the approach with two use cases.

\end{abstract}
\section{Introduction}
\label{sec:intro}
\begin{figure}
\centering
\includegraphics[width=0.7\linewidth]{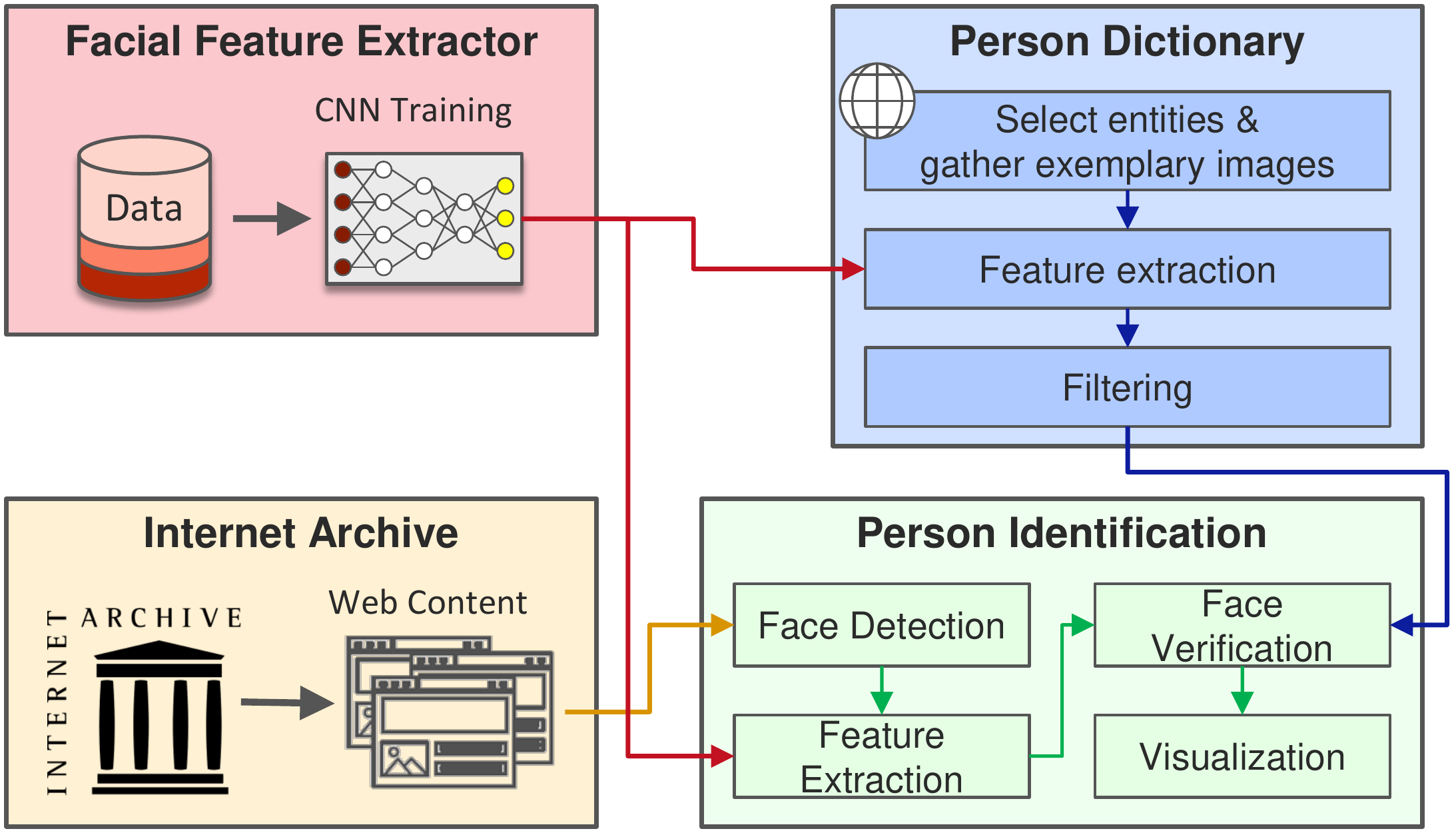}
\caption{Workflow of the proposed person identification framework}
\label{fig:workflow}
\end{figure}
The World Wide Web with its billions of web pages and the respective multimedia content includes valuable information for many academic and non-academic applications. Therefore, national (digital) libraries and the \ia~ (\url{www.archive.org}) have been capturing the available multimedia data in huge web archives with time-stamped snapshots of the web pages since the mid-nineties. This serves as a playground for researchers and analysts in different domains such as politics, economics, and entertainment. One of the main challenges is to make the available unstructured data, which is rarely enriched with appropriate metadata, accessible and explorable by the users. 
For this reason, it is necessary to develop (semi)-automatic content analysis 
approaches and systems to automatically extract metadata that can be subsequently used for 
semantic search and information visualization in order to provide users with relevant information 
about a given topic. 
In recent years, many tools like \emph{BabelNet}~\cite{navigli2012babelnet}, 
\emph{Dandelion}~\cite{brambilla2017extracting}, and 
\emph{FRED}~\cite{gangemi2017semantic} have been introduced that aim to track 
entities and their relations using textual information. 
However, we argue that text does not cover every entity in general and that image (and video) data can contain additional information. Visual and textual content can be complementary and their combination can serve as a basis for a more complete entity recognition system. However, approaches that exploit image or video data in the \ia~are rare. 

In this paper, we present a system (Figure~\ref{fig:workflow}) that enables researchers and analysts to find and explore
media coverage and relations of persons of interest in the image content of the Internet Archive for a given domain such as politics, sports and entertainment. A number of sample images is crawled for each entity using an image search engine such as \emph{Google Images}. Due to noise 
in the retrieved data of this web-based approach, we investigate two strategies 
to improve the quality of the sample dataset. A state of the art convolutional neural network~(CNN) is used to learn a robust feature representation in order to describe the facial characteristics of the 
entities. Afterwards, the trained deep learning model is applied based on the sample dataset to identify the selected entities in the image content of all German web pages in the \ia. All required processing steps in the 
pipeline are designed to match the requirements of this big data application in terms of the 
computational efficiency. The performance of our CNN-based feature representation is evaluated on the 
Labeled Faces in the Wild~(LFW) dataset~\cite{huang2007labeled}. Finally, we evaluate the performance of our automatic web-based system by presenting two use cases along with appropriate graphical representations.
To the best of our knowledge, this is the first approach to identify entities in 
the \ia~using solely image data.  

The remainder of the paper is organized as follows. A brief overview of related work for face recognition and 
entity extraction is given in Section~\ref{sec:rw}. In Section~\ref{sec:person_identification}, we introduce 
our deep learning system to identify persons in the \ia. Experimental results for the face recognition approach as 
well as some use cases are presented in Section~\ref{sec:exp}. The paper concludes in Section~\ref{sec:conc} with a 
summary and areas of future work.
\section{Related Work}
\label{sec:rw}
In recent years a number of very powerful tools performing named entity 
recognition like 
\emph{BabelNet}~\cite{moro2014entity,navigli2012babelnet}, 
\emph{Dandelion}~\cite{brambilla2017extracting}, 
\emph{FRED}~\cite{gangemi2017semantic} and \emph{NERD}~\cite{van2013learning} 
have been developed. In particular, the recognition of entities depicting public 
personalities already achieves very good results. But especially online news 
articles are often provided with photos, which potentially show additional entities 
that are not mentioned in the text. Furthermore, possible disambiguations could 
be resolved using the visual content. 
For this reason, face recognition approaches can be applied to predict persons 
in the images. This is a well studied computer vision task for decades and the 
performance has been significantly improved  since convolutional neural 
networks~\cite{krizhevsky2012imagenet} and huge public data collections like 
\emph{CASIA-WebFace}~\cite{yi2014learning} and 
Microsoft-Celebrity-1M~(\emph{MS-Celeb-1M})~\cite{MS-Celeb-1M} have been introduced. 
As one of the first CNN-based approaches DeepFace~\cite{taigman2014deepface} 
treats face recognition as a classification approach and subsequently uses the 
learned feature representation for face verification. 
In general, face recognition approaches based on deep learning benefit 
from learning a robust face representation, new loss functions like the 
contrastive loss~\cite{sun2014deep}, triplet loss~\cite{schroff2015facenet} and 
angular softmax~\cite{liu2017sphereface} have been introduced to enhance the 
discriminative power. 
To improve the robustness against pose variation some 
approaches~\cite{zhu2016face,masi2016pose,masi2016we,yin2017towards,masi2017rapid,masi2018learning} aim to frontalize the face using 3D-head models or synthesize new views of the 
face to augment the training dataset with all available poses, respectively.
Another widely used technique to increase the robustness to poses and occlusions are approaches~\cite{yang2015from,ding2017trunk} that use several image patches around facial landmarks as input for the CNN network training. 
To overcome variations due to the aging of faces, several approaches~\cite{wen2016latent,best2018longitudinal} are suggested as well. 
\section{Person Identification in the Internet Archive}
\label{sec:person_identification}

In this section, a system for the identification of interesting persons in archived web web news is introduced. First, a CNN is trained to learn a robust representation for faces (Section~\ref{sec:feature_extractor}). In Section 3.2, we describe a way to  define a lexicon of persons and to automatically gather sample images from the Web for them, i.e., to build an entity dictionary for a given domain like politics or entertainment. In this context, we explain how to reduce noise in the sample dataset which is due to the web-based approach. The proposed framework retrieves image data from the \ia~according to a predefined search space (Section~\ref{sec:ia_image_retrieval}). 
Section~\ref{sec:person_identification_workflow} presents the complete workflow for identifying persons in images of the \ia~based on the predefined dictionary. Furthermore, single and joint occurrences of persons in images are explored and visualized. The workflow of the approach is illustrated in Figure~\ref{fig:workflow}.
\subsection{Learning a Feature Representation for Faces}
\label{sec:feature_extractor}
To learn a reliable representation for person identification in the subsequent steps, a CNN is trained. Given a dataset of face images such as \msc~\cite{MS-Celeb-1M} or \emph{CASIA-WebFace}~\cite{yi2014learning} covering $n$ individual persons, a model with the number of classes $n$ is trained for classification. During training the cross-entropy loss is minimized given the probability distribution $C$ for the output neurons and the one-hot vector $\hat{C}$ for the ground-truth class:
\begin{equation}
E(C,\hat{C}) = - \sum_i \hat{C_i} log(C_i).
\end{equation}
Removing the fully-connected layer that assigns probabilities to the predefined classes of faces transforms the model to a generalized feature extractor. Thus, for a query image the model outputs a compact vector of facial features. In this way a query image can be compared with the facial features of entities in the predefined dictionary, which is presented in the next section.
\subsection{Creating a Dictionary of Persons for a Domain}
\label{sec:person_dict}
First, the web-based steps required to automatically define entities and gather sample images for them are explained. Afterwards, the process of defining a compact representation for each entity is described. In this context, two strategies for filtering inappropriate facial features are introduced and discussed. 
\subsubsection{Defining Entities and Gathering Sample Images:}
In order to automatically define relevant persons of a given domain  of interest, a knowledge base 
such as the \emph{Wikipedia} encyclopedia is queried for persons associated with the target group, e.g., politicians. To retrieve the most relevant entities for the selected group, the query is further constrained to persons whose pages were viewed most frequently in a given year and who were born after 1920. 
Sample images for the selected persons are retrieved in an unsupervised manner employing a web-based approach. Given the names of the selected entities, an image search engine such as \emph{Google Images} is crawled to find a given number of $k$ sample images for each person. However, the collected images do not necessarily always or only depict the target person but involve some level of noise which should be eliminated in the following steps.
\subsubsection{Extraction and Filtering of Feature Vectors:}
To extract face regions in an image, we use \emph{dlib} face detector based on the \emph{histogram of oriented gradients}~(HOG)~\cite{dalal2005histograms}. This face detector ensures efficiency in terms of computational speed, especially for the detection of faces in the large-scale image data of the \ia~ (see Section~\ref{sec:person_identification_workflow}). For each detected face $i$ associated to person $p$ the feature vector $f_i \in F_p, i= 1,\dots, |F_p|$ is computed using the CNN model ( Section~\ref{sec:feature_extractor}). Since the detected faces can depict the target person as well as other individuals in the corresponding image or entirely different persons due to noise, a data cleansing step on the extracted facial features $F_p$ is conducted. In particular, the cosine similarity
of each feature vector $f_i$ to a target feature vector $f_T$ representing the individual $p$ is computed. For the choice of the target feature vector, we propose (1) using the mean of all feature vectors in $F_p$, or (2) selecting one example vector among those in $F_p$. Facial feature vectors yielding a similarity value smaller than a given threshold $\lambda_1$ are removed. Choosing the mean feature vector is advantageous in the sense that it does not require supervision, unlike the manual selection of exemplary vectors for the dictionary entries. On the other hand, the average vector may not be meaningful if the images collected from the web contain a lot of noise. In addition, the selection of exemplary face vectors in a supervised manner unambiguously represents the target entities and thus ensures a more robust filtering of false positives. The evaluation of the proposed vector choices for filtering as well as the choice of threshold $\lambda_1$ are discussed in Section~\ref{sec:exp}. 
\subsubsection{Definition of the Final Dictionary:}
After the filtering step is applied, the set of the remaining facial features $F_p$ is represented by the mean vector:
\begin{equation}
\overline{f_p} = \frac{1}{|F_p|} \sum_{i=1}^{|F_p|} f_i.
\end{equation}
This is computationally advantageous for the subsequent steps in terms of numbers of comparisons. For the resulting dictionary $D_P = \{\overline{f_1}, \ \dots, \ \overline{f_{|P|}}\}$ only one computation per person is required to verify if a query depicts the given entity, instead of comparing against the entire set of feature vectors describing an individual entity.
\subsection{Retrieving Images from the Internet Archive}
\label{sec:ia_image_retrieval}
The \ia~contains an enormous amount of multimedia data that can be used to reveal dependencies between entities in various fields. Looking only at the collection of web pages, a large part of the multimedia content is irrelevant for person search, e.g., shopping websites. For this reason, we aim at selecting useful and interesting domains in which the entities from the dictionary are depicted. In particular, we retrieve image data of web pages such as the German domain \url{welt.de} for political subjects. Furthermore, the amount of images is restricted to valid image formats as JPEG and PNG, excluding formats like GIF can reduce the amount of possible spam. Another useful criterion for restricting the image search space is the publication date, which enables the exploration of events of a certain year.
\subsection{Person Identification Pipeline}
\label{sec:person_identification_workflow}
Using the previously introduced components, persons in the image data of the \ia~can be automatically identified.
With the aid of a HOG-based face detector~\cite{dalal2005histograms}, first, all available faces in the set of retrieved images of the \ia~are extracted. Based on the CNN described in Section~\ref{sec:feature_extractor}, feature vectors for the faces are computed. Subsequently, each query vector is compared against the dictionary of persons (Section~\ref{sec:person_dict}). 
The cosine similarity of a query vector to each entity vector is computed, in order to determine the most similar person in the dictionary. Given the similarity value, a threshold $\lambda_2$ verifies whether the potential dictionary entity is depicted. 
Based on the results of person identification, visualizations based on single and joint occurrences of persons of interest in the selected \ia~data can be created.
\section{Experimental Setup and Results}
In this section, we evaluate the components of our deep learning-based person identification framework. We present details of the technical realization as well as experimental results on the learned feature representation for faces (Section \ref{sec:exp_cnn}) and the dictionary of persons (Section \ref{sec:exp_dict}). We select two example groups of prominent people. The feasibility of our system is demonstrated on image data of the \ia~concentrating on a selection of German web content (Section \ref{sec:exp_pipeline}). Finally, visualizations for relations among the persons of interest in the selected data are shown.
\label{sec:exp}
\subsection{CNN for Facial Feature Extraction}
\label{sec:exp_cnn}
\subsubsection{Training of the CNN Model:}
Several publicly available datasets exist for the person recognition task. We use the face images of MS-Celeb-1M~\cite{MS-Celeb-1M} as input data for our CNN. Comprising $8.5\,M$ images of around $100\,K$ different persons, it is the largest publicly available dataset. A classification model considering all the available identities of the dataset is trained using the \emph{ResNet} architecture~\cite{he2016deep} with 101 convolutional layers. The weights are initialized by a pre-trained ImageNet model.
Furthermore, we augment the data by randomly selecting an area covering at least $70\%$ of the image. The input images are then randomly cropped to $224 \times 224$ pixels.
Stochastic Gradient Descend (SGD) is used with a momentum of 0.9. The initial learning rate of 0.01 is exponentially decreased by a factor of 0.5 after every 100,000 iterations. The model is trained for 500,000 iterations with a batch size of 64. The training was conducted on two Nvidia Titan X graphics cards with 12 GB VRAM each. The implementation is realized using the TensorFlow library~\cite{abadi2016tensorflow} in Python. The trained model is available at: \url{https://github.com/TIB-Visual-Analytics/PIIA}.
\subsubsection{Evaluating the CNN Model:}
The trained model is evaluated on the well-known and challenging LFW benchmarking set~\cite{huang2007labeled} in the verification task. Therefore, the similarity between two feature vectors using the cosine distance is measured.
We perform 10-fold cross validation experiments, where each fold consists of respectively 300 matched and mismatched face pairs of the test set. For each subset, the best threshold maximizing the accuracy on the rest of 9 subsets is calculated. Thus, the yielded accuracy as well as threshold values are averaged for the 10 folds.
The trained model obtains an accuracy of $98.0\,\%$ with a threshold set to $0.757$. 
Compared to the much more complex systems achieving state of the art results on the LFW benchmark, our model yields a satisfactory accuracy using a base architecture and loss function and provides a good basis for the face verification step in our pipeline. Moreover, the estimated threshold has a standard deviation of 0.002. This indicates that the threshold value is very stable for the variety of input faces. In the following, the threshold for comparing a target vector against dictionary entity vectors for filtering (Section \ref{sec:person_dict}) is assigned to $\lambda_1=0.757$. 
\subsection{Creating a Dictionary of Persons}
\label{sec:exp_dict}
\subsubsection{Selecting Entities and Gathering Image Samples:}
In our experimental setting, the goal is to recognize persons in the German web content of the \ia~and visually infer relations among them. Hence, people of public interest have to be selected for the dictionary. We exemplarily choose the groups of \emph{politicians} and \emph{actors}, for each of whom we create a dictionary according to the description in Section \ref{sec:person_dict}.  
To obtain international as well as German personalities, especially for the case of politicians, we query the German \emph{Wikipedia} for persons according to the selected occupations and further criteria specified in Section \ref{sec:person_dict}. The entity names are fetched via SPARQL queries to the Wikidata repository, along with the number of page views. Since Wikidata provides page views from mid 2015, we fetch the numbers for the year 2016. This results in a minor mismatch in terms of time concerning our search space for the \ia~data which we discuss and clarify in Section~\ref{sec:exp_pipeline}. 
The relevance of the collected entities is subsequently determined by the number of page views. Thus, the ranked list of entity names is reduced to the first 100 entries. Given the sets of persons for the selected occupational groups, we crawl the \emph{Google Images} search engine for a maximum of 100 images per entity.
\begin{table}[b]
\caption{Results of methods for the cleansing step of the entity dictionary on a subset of 20 politicians.}

\centering
	\begin{tabular}{L{2.5cm} C{2.0cm} C{2.0cm} C{2.0cm}}
    	\textbf{Method} & \textit{Precision} & \textit{Recall} & \textit{$F_1$} \\ \hline \hline
        no filtering & 0.669 & 1 & 0.802 \\ \hline
		mean vector  & 0.993  & 0.449 & 0.618\\ \hline
        reference vector & 0.977  & 0.922 & 0.949 \\
    \end{tabular}
\label{tab:filtering_results}
\end{table}

%
\subsubsection{Evaluating the Methods for Feature Vector Cleansing:}
\begin{figure}[hb]
\centering
\includegraphics[width=0.95\linewidth]{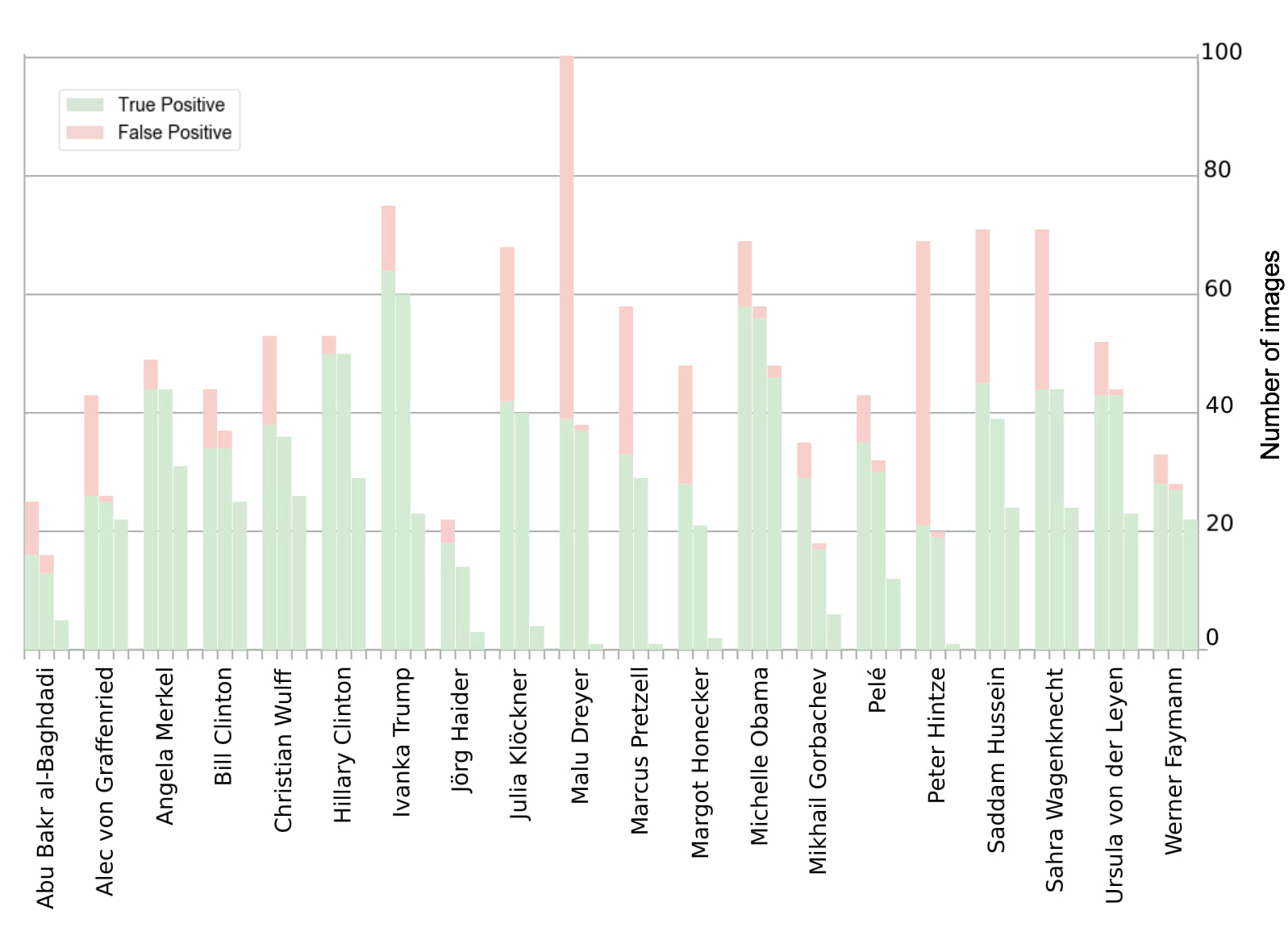}
\caption{Composition of true and false positives according to filtering methods on given entity test set. Grouped bars denote (a) unfiltered entity images, (b) filtering with a reference vector and (c) filtering according to the mean vector.}
\label{fig:barchart_filtering}
\end{figure}

In Section \ref{sec:person_dict} two methods for selecting a target vector for filtering entity vectors are introduced. Using the threshold $\lambda_1$ which was estimated on the LFW benchmark, we separately filter the entity vectors according to the average entity vector and a manually chosen reference vector. The methods are evaluated on an annotated subset of 1100 facial images covering 20 dictionary entities of the group of politicians. 
Table~\ref{tab:filtering_results} reports filtering results for both strategies in comparison to the unfiltered test set. As shown, the use of the mean vector boosts precision from initially 0.669 to 0.993, but recall is reduced to 0.449. As already hypothesized in Section \ref{sec:person_dict}, this is due to the noise in the exemplary images caused by the web-based approach and thus the strong distortion of the mean vector. 
Figure~\ref{fig:barchart_filtering} illustrates that the average vector drastically discards false as well as true positives making this strategy impractical for our purposes. In comparison, the manually selected vector boosts recall to 0.922 and thus significantly reduces the false positive rate of entity images (see also Figure~\ref{fig:barchart_filtering}). For most depicted entities such as \emph{Malu Dreyer} or \emph{Angela Merkel} almost every false positive is filtered out while the correct images are maintained. The method yields a slightly smaller precision of 0.977 and requires supervision which we take into account for the subsequent steps due to the high $F_1$ score.
\subsubsection{Evaluating a Global Threshold for Face Verification:}
After noisy vectors are filtered out, each entity is described by its mean vector. The use of a mean vector is plausible for our framework since we do not detect and identify faces in extreme poses. Since the global face verification step for query images is carried out with the average of all vectors of a entity, a new threshold $\lambda_2$ value different than $\lambda_1$ has to be estimated. For this reason, a cross-fold validation in the same way as in Section~\ref{sec:exp_cnn} is performed based on the subset of politicians used for the evaluation of the dictionary cleansing. An accuracy of $96\%$ is obtained. The threshold results in $\lambda_2=0.833$ and shows a standard deviation of 0.002. In particular, the very small standard deviation implies that the use of the mean entity vector works very stable for the face verification task of our framework.
\subsection{Face Recognition in Image Collections of the \ia~}
\label{sec:exp_pipeline}
\subsubsection{Selection of Image Data:}
To demonstrate our person identification framework upon selected web content of the \ia, we select the two German news websites \url{welt.de} and \url{bild.de}. While the former addresses political subjects, \url{bild.de} mostly focuses on entertainment news as well as celebrity gossip. Therefore, we separately exploit \url{welt.de} for identifying politicians and \url{bild.de} for identifying the selected actors. We select image data of the year 2013, in which the German elections took place. Please not the minor offset compared to the selected entities using statistics of \emph{Wikipedia} from 2016~(Section~\ref{sec:exp_dict}). However, the persons are identifiable and relevant.
\begin{figure}[h]
\centering
\includegraphics[width=0.8\linewidth]{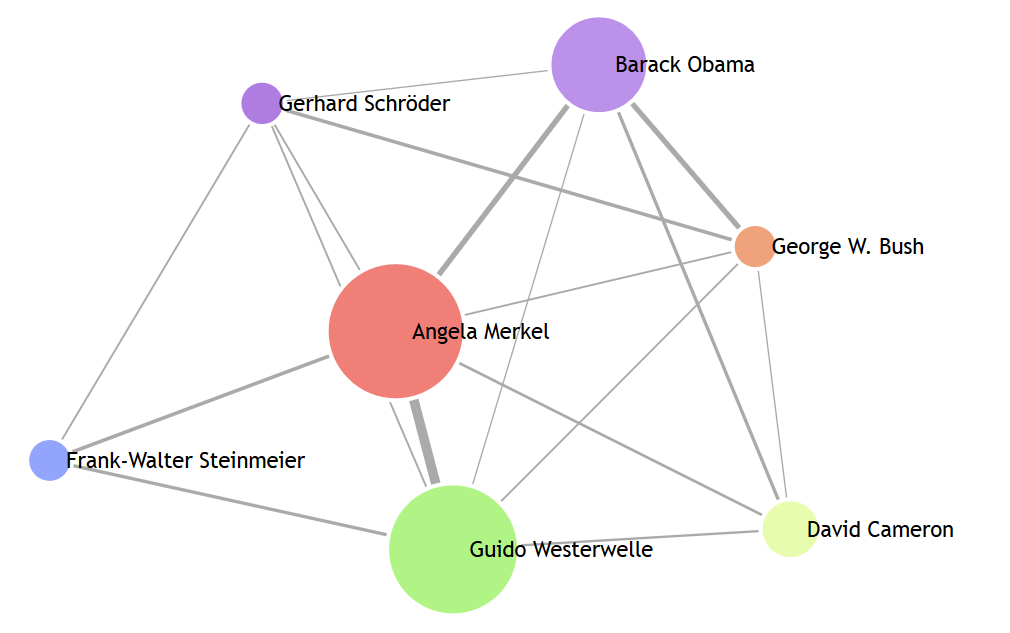}
\caption{Graph showing relations among an exemplary set of politicians. The size of vertices encodes the occurrence frequency of the entity. The strength of edges denotes the frequency of joint occurrences.}
\label{fig:graph_politicians}
\end{figure}
\subsubsection{Visualization:}
To quantify the relevance of individuals and their relation to other entities, we count how often single entities appear in the selected image data and how often they are portrayed with persons of the dictionary. 
Figure~\ref{fig:graph_politicians} visualizes relations between well-known heads of states and other politicians in 2013 inferred by our visual analysis system for the German news website \url{welt.de}. The graph shows that \emph{Angela Merkel}, the German chancellor, as well as the former German minister of foreign affairs, \emph{Guido Westerwelle}, appear most frequently in the image data and also share a strong connection. The most relevant international politician detected in the news images is \emph{Barack Obama} with a strong connection to \emph{Angela Merkel}. The connection of \emph{Westerwelle} to \emph{Steinmeier} is due to the transition of power in late 2013. Also, connections between former and new heads of states of Germany and the USA exist. 
\begin{figure}[t]
\centering
\includegraphics[width=0.9\linewidth]{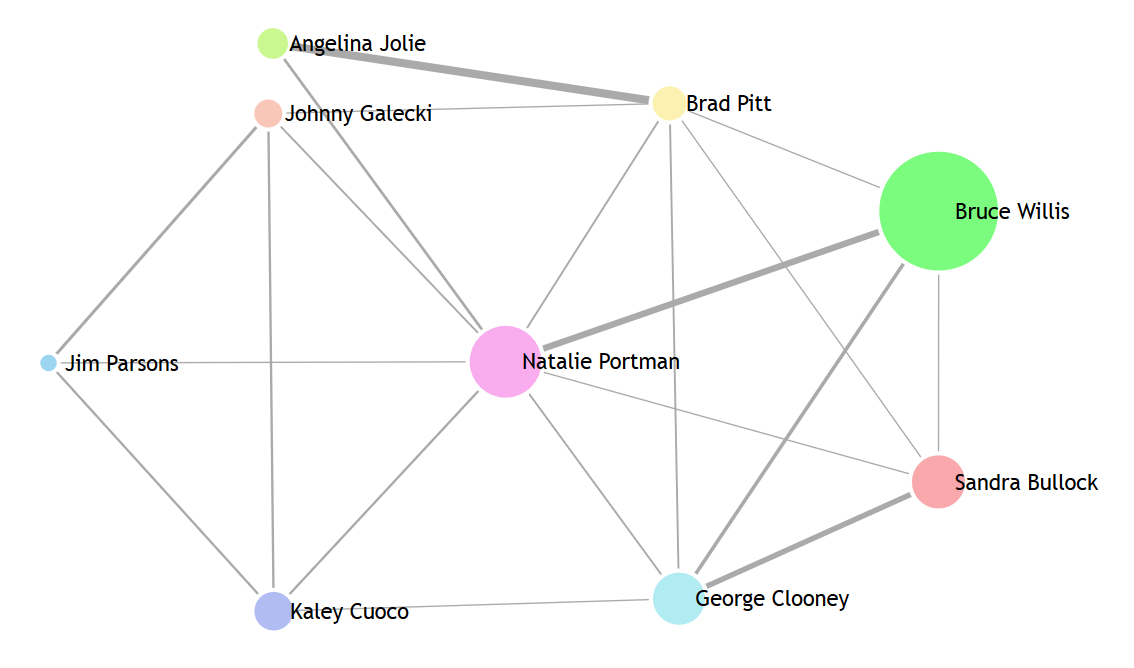}
\caption{Graph showing relations among an exemplary set of actors. The size of vertices encodes the occurrence frequency. The strength of edges denotes the frequency of joint occurrences.}
\label{fig:graph_actors}
\end{figure}
\newline In the same manner, Figure~\ref{fig:graph_actors} visualizes connections between different actors in image data of \url{bild.de} published in 2013. For example, the graph reveals strong connections between the actors \emph{George Clooney} and \emph{Sandra Bullock} who have acted in the same movie in 2013. Moreover, actors of a well-known sitcom share connections with each other. Also a strong connection between an actor couple in 2013 can be determined. The actress \emph{Natalie Portman} provides connections to all actors of the graph having the second strongest appearance frequency. This implies that there must be several images published in \url{bild.de} which depict the actress with her actor colleagues, maybe due to a celebrity event like the Academy Awards. 
\section{Conclusions}
\label{sec:conc}
In this paper, we have presented an automatic system for the identification of persons of interest in image content of web news in the \ia. 
For this task, a CNN-based feature representation for faces was trained and evaluated on the standard LFW benchmark set. Moreover, we introduced a semi-automatic web-based method for creating a dictionary of persons of interest, given an area of interest. In addition, two methods for filtering inappropriate images in the sample data were introduced and evaluated. In order to cope with the enormous amount of image content the \ia~provides, a constrained search domain was defined. The proposed system reliably detects dictionary entities and reveals relations between the entities by means of joint occurrences. In order to process the huge amount of image data, the system is realized using efficient and well scalable solutions. 

In the future, we plan to further improve individual steps of the pipeline. In particular, we aim to improve our deep learning model using a more sophisticated loss function like the triplet loss~\cite{schroff2015facenet} or preprocessing for more robustness against pose variation. In addition, a face detector will be used that deals with arbitrary poses. The process of determining a ground truth vector can be automated by querying \emph{Wikipedia} for a representative image of the entity. Finally, the framework will be extended to allow the exploration of relations of persons across different domains. 

\section*{Acknowledgement}
This work is financially supported by the German Research Foundation (DFG: Deutsche Forschungsgemeinschaft, project number: EW 134/4-1). The work was partially funded by the European Commission for
the ERC Advanced Grant ALEXANDRIA (No. 339233, Wolfgang Nejdl).

\bibliographystyle{splncs04}
\bibliography{piia}
\end{document}